\documentclass[prb,twocolumn,showpacs,preprintnumbers,amssymb]{revtex4}
\usepackage{graphicx}
\usepackage[tbtags]{amsmath}
\usepackage{bm}

\begin{document}

\title
{Cooling a magnetic resonance force microscope via the dynamical
back-action of nuclear spins}
\author{Ya.~S.~Greenberg$^{\rm 1,2}$, E.~Il'ichev$^{\rm 3}$, and Franco Nori$^{\rm 1,4}$}

\affiliation{$^{1}$Advanced Science Institute, The Institute of
Physical and Chemical Research (RIKEN), Wako-shi, Saitama
351-0198, Japan}

\affiliation {$^{2}$Novosibirsk State Technical University, 20
Karl Marx Ave., 630092 Russia}

\affiliation{$^{3}$Institute of Photonic Technology, PO Box
100239, D-07702 Jena, Germany,}

\affiliation{$^{4}$Department of Physics, Center for Theoretical
Physics, and Center for the Study of Complex Systems, University
of Michigan, Ann Arbor, MI 48109-1040, USA}

\date{\today}
\begin{abstract}
We analyze the back-action influence of nuclear spins on the
motion of the cantilever of  a magnetic force resonance
microscope. We calculate the contribution of nuclear spins to the
damping and frequency shift of the cantilever. We show that, at
the Rabi frequency, the energy exchange between the cantilever and
the spin system cools or heats the cantilever depending on the
sign of the high-frequency detuning. We also show that the spin
noise leads to a significant damping of the cantilever motion.

\end{abstract}

\maketitle
\section{Introduction}
\label{intr}

Magnetic resonance force microscopy (MRFM) is a powerful technique
for visualizing subsurface
structures\cite{Sidles,Suter,Ku,Ham,Ham1} with three dimensional
spatial resolution of the order of 10 nanometers or less
\cite{Deg}, which is more than two orders of magnitude better than
the resolution of conventional high-field magnetic resonance
imaging (MRI). The main part of a MRFM device is a nanomechanical
resonator or cantilever (see Fig.~\!\ref{cantilever} below),
having a fundamental frequency in the range of several kHz. By
using MRFM, a significant breakthrough in magnetic resonance
detection sensitivity was achieved, resulting in single-electron
spin detection\cite{Rugar} with a spatial resolution $\sim$25 nm
and substantial progress in nuclear spin
detection\cite{Rug,Kent,Kent1,Mamin,Mam}. MRFM has also been
proposed as a qubit readout device for spin-based quantum
computers\cite{Berman1,Pel}.

MRFM was initially proposed as a possible means to improve the
detection sensitivity to the single spin level\cite{Sidles}. Since
then, progress in MRFM and related technologies has attracted
broad interest, especially the questions of squeezed states of the
cantilever and the collapse of its wave function when both, the
spin to be measured and the cantilever, are treated quantum
mechanically \cite{Xue1,Xue2,Gassmann, Berman2, Berman3}. However,
the ultimate goal to detect a single nuclear spin with MRFM is
still a challenge. Rough estimates indicate that in order to reach
this goal the effective temperature of the MRFM cantilever should
be reduced to about 0.1 $\mu$K, which corresponds to $\sim$2.5 kHz
of the fundamental frequency of the MRFM cantilever.

Numerous experiments on cooling micro-mechanical resonators via
their coupling with different external systems have recently been
reported (see, e.g., Ref.~\onlinecite{Kleck,Met,Teuf,Pog,Schlis}).
Experimental results show that a micro-mechanical resonator can be
cooled down to an effective temperature on the order of 0.1
K(Ref.~\onlinecite{Kleck}) or 5 mK (Ref.~\onlinecite{Pog}).
However, in order to drive the micro-resonator to the quantum
regime, more effective cooling methods are needed. A promising way
would be to cool the micro-resonator by coupling it to a
solid-state quantum electronic circuit. In principle, the
effective electronic cooling of the micro-resonator can be
achieved by several means, including coupling it to an another
resonator\cite{Graj,Brown}, to a transmission line
resonator\cite{Zhang,Xue}, to a quantum dot\cite{Ouy}, to an
electronic spin\cite{Rabl}, or to superconducting devices
\cite{Ilich,Nori,Grajcar,Val,Hau,Hau1,Green1,Green2,You,Naik,Martin,Blenc,Clerk}
.

In particular (as shown, e.g., in
Refs.~\onlinecite{Ilich,Grajcar,You1,Ash,Hau,Hau1,Green1,Green2}),
an electric resonator circuit weakly coupled to a two-level system
(superconducting flux qubit) can be \emph{cooled} by its
quantum-dynamical \emph{back-action}.

In this paper, we investigate the cooling of a MRFM cantilever via
its coupling to nuclear spins. We show that the \emph{back-action}
of the spin system modifies the equation of motion of the
cantilever, providing  additional damping and a frequency shift
which depends on the properties of the spin system (decoherence
rates, damping rates, etc). We investigate the operation modes of
the MRFM where the damping is positive and results in a
substantial decrease of the effective quality factor of the MRFM
cantilever, and thus a significant \emph{cooling} of the
cantilever motion.

This paper is organized as follows. In Section \ref{interaction}
we describe the interaction of a MRFM cantilever with nuclear
spins. We obtain the cantilever  equation of motion modified by
the back-action of the spin system. The modification appears as an
additional contribution to the damping and frequency shift of the
cantilever, in terms of the magnetic spin susceptibility.

In Section \ref{lfsusc} we obtain the explicit expression for the
low-frequency spin susceptibility. The most important result of
this section is that the \emph{longitudinal} magnetization of a
sample has a clear resonance at its Rabi frequency. In some sense,
this is the \emph{low-frequency analog of the conventional
high-frequency} NMR for the \emph{transverse} magnetization.

Section \ref{influence} is devoted to a detailed study of the
influence of the spin system on the damping and the frequency
shift of the cantilever. In the first part of this section we
consider a sample with a relative short spin-lattice relaxation
time $T_1$. In this case, we show that the quality factor of the
cantilever changes depending on the sign of the high-frequency
detuning. For positive detuning (the microwave frequency is above
the nuclear resonance frequency), the contribution of the spin
system to the cantilever damping is negative: heating the
cantilever by absorbing Rabi photons from the spin system. If the
detuning is negative (the microwave frequency is below the nuclear
resonance frequency), the contribution of the spin system to the
cantilever damping is positive: \emph{cooling the cantilever by
giving up Rabi photons to the spin system}. In the second part of
Section \ref{influence} we consider the influence of the spin
noise on the cantilever motion. Independently of the particular
values of the parameters which characterize the spin system, we
show that its influence on the cantilever damping is always
positive, i.e., the spin noise always leads to a decrease of the
quality factor of the cantilever.

\section{Interaction of nuclear spins with the MRFM cantilever}
\label{interaction}

\begin{figure}
\includegraphics[width=\columnwidth]{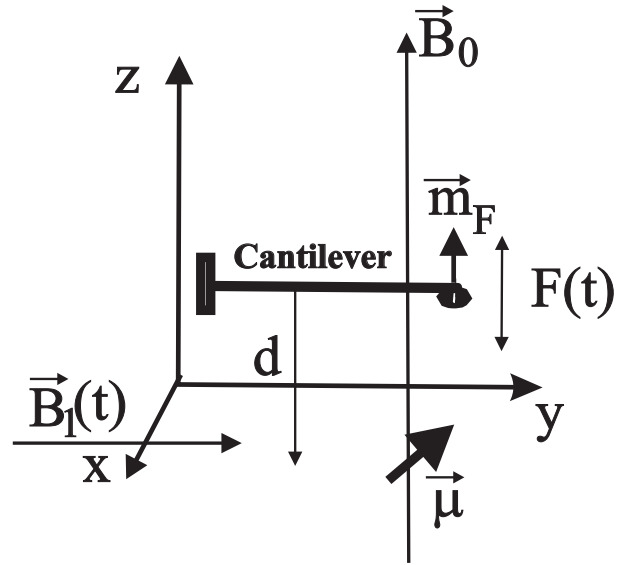}
\caption{\small Schematic diagram of the setup of the system under
consideration. Here, $B_0$ is the uniform permanent magnetic
field, $B_1(t)$ is the rotating rf magnetic field, $F(t)$ is an
external force acting on the cantilever in the $z$-direction,
$m_F$ is the magnetic moment of the ferromagnetic particle
attached to a free end of the cantilever, $\mu$ is the magnetic
moment of the nuclear spin, located in $xy$- sample plane, and $d$
is the equilibrium distance between the center of the
ferromagnetic particle and a sample plane.}\label{cantilever}
\end{figure}

A schematic diagram of the  system studied here is shown in
Fig.~\ref{cantilever}.
A spherical ferromagnetic particle with magnetic moment $m_F$ is
attached to the cantilever tip. A small paramagnetic cluster with
magnetic moment $\mu$, which must be detected, is placed on the
surface of a non-magnetic sample beneath the tip of the
cantilever. The whole system is placed in a permanent high
magnetic field, $B_0$, oriented in the positive $z$-direction. The
transverse magnetic field $B_1(t)$, which excites the NMR in the
sample, is applied to the paramagnetic cluster. In addition to
$B_0$, the magnetic moment $\mu$ experiences the inhomogeneous
field $B_F(z)$ from the ferromagnetic tip. We assume that the
field $B_F(z)$ is also oriented in the positive $z$-direction and
is given by the dipole formula:
\begin{equation}\label{Bf}
    B_F(z)=\frac{\mu_0}{4\pi}\frac{2m_F}{(d+z)^3}\; ,
\end{equation}
where $d$ is the equilibrium distance between the cantilever and
the sample surface, and $z$ is the amplitude of the cantilever
oscillations. Below we assume $z\ll d$, hence

\begin{equation}\label{B1}
    B_F(z)\approx
    B_F(0)\left(1-\frac{3}{d}z+\frac{6}{d^2}z^2\right)\; ,
\end{equation}
where
$$
 B_F(0)=\frac{2\mu_0 m_F}{4\pi d^3}\ \ .
$$
Hence, the cluster we investigate is under the polarizing field:
$B_0+B_F(0)$.

The interaction of a particle, having a magnetic moment $\mu$,
with the cantilever is given by the following Hamiltonian
\begin{equation}\label{H}
    H=H_C+H_C^{(b)}+H_{CB}+H_S+H_S^{(b)}+H_{SB}-
    (\overrightarrow{\mu}\cdot \overrightarrow{B}),
\end{equation}
where $H_C$ is the Hamiltonian of the cantilever
\begin{equation}\label{HC}
    H_C=\frac{p^2}{2m}+\frac{m\omega^2z^2}{2}\; ,
\end{equation}
$H_S$ is the Hamiltonian of a nuclear spin interacting with a
one-mode high-frequency field
\begin{equation}\label{HS}
    H_S=\frac{\hbar\omega_0}{2}\sigma_Z+
    \frac{\hbar\omega_1}{2}\sigma_X(a^++a)
+\hbar\omega_{\rm mw}a^+a\; ,
\end{equation}
where
$$
\omega_0=\gamma (B_0+B_F(0))\ ,
$$
$\gamma$ is the nuclear gyromagnetic ratio, $\omega_{\rm mw}$ is
the frequency of the microwave field. The quantity
$\hbar\omega_1$, the interaction energy between spin and the
microwave field, is proportional to its amplitude $B_1$ (see
below). The magnetic moment $\overrightarrow{\mu}$ of a spin-$1/2$
particle is expressed in terms of the Pauli spin matrix vector:
$\overrightarrow{\mu}=-\hbar\gamma\overrightarrow{\sigma}/2$.

The Hamiltonians $H_C^{(b)}$ and $H_S^{(b)}$ represent the thermal
baths for the cantilever and spin, respectively, while $H_{CB}$
and $H_{SB}$ represent their interactions with their corresponding
baths.

We will not specify here the bath Hamiltonians $H_C^{(b)}$,
$H_S^{(b)}$, and their interactions $H_{CB}$, $H_{SB}$. We
describe the influence of $H_{C}$, $H_{CB}$ on the motion of the
cantilever by introducing the damping rate $\gamma_c$ and the
external noise $\eta(t)$.

We consider the cantilever tip as an oscillator with effective
mass $m$, and effective spring constant $k_s$, subject to an
external force $F_0\cos\omega t$ and force fluctuations with a
spectral density
\begin{equation}\label{forcenoise}
S_F(\omega)=2\hbar\omega\frac{ m\omega_c}{Q_c}
\coth\left(\frac{\hbar\omega}{2kT}\right)\ ,
\end{equation}
where $Q_c$ is the quality factor of the bare cantilever.

The equation of motion for the cantilever, interacting with $N$
spin-$1/2$ particles, then reads:
\begin{equation}\label{cant}
    \ddot z + \gamma _c \dot z + \omega _c^2 z =
     \left\langle\frac{\widehat{M}_Z}{m}\frac{{dB_F}}{{dz}}\right\rangle+f_0\cos\nu t+{\eta}(t),
\end{equation}
where $\widehat{M}_Z$ is the quantum operator of the longitudinal
magnetization,
$$
\widehat{M}_Z=-\frac{N\hbar\gamma}{2}\sigma_Z\; ,
$$
$\omega_c=(k_s/m)^{1/2}$, $\gamma_c=\omega_c/Q_c$, $f_0=F_0/m$,
and $\eta(t)$ describes the fluctuations of the acceleration, and
has a spectral density $S_{\eta}(\omega)=S_F(\omega)/m^2$.

 By using Eq. \!(\ref{B1}) we obtain:
\begin{equation}\label{inter}
    \left\langle\frac{\widehat{M}_Z}{m}\frac{{dB_F}}{{dz}}\right\rangle=\lambda
    \langle\sigma_Z\rangle-\frac{4\lambda}{d}\langle z\sigma_Z
    \rangle\; ,
\end{equation}
where
$$
\lambda=\frac{{3N\hbar\gamma}B_F(0)}{2md}
$$
is the coupling strength between the spin and the cantilever. The
angular brackets in (\ref{cant}) and (\ref{inter}) denote the
average over the two free baths variables.

The quantity $\langle\sigma_Z\rangle$ in (\ref{inter}) is a
functional of the cantilever position $z(t)$, where its first
order ($\approx z/d$) contribution to the Hamiltonian (\ref{H}) is
$(\lambda m z+f)\sigma_Z$, where we introduce $f(t)$: a small
external force that is required for calculating the magnetic
susceptibility\cite{Smirnov}. Hence, to first order in $\lambda$,
we obtain
\begin{equation}\label{dir}
    \langle\sigma_Z(t)\rangle=\langle\sigma_Z^{(0)}(t)\rangle+\lambda m\int
    dt_1\frac{\delta\langle\sigma_Z(t)\rangle}{\delta f(t_1)}
    z(t_1)\; ,
\end{equation}
where $\langle\sigma_Z^{(0)}(t)\rangle$ is described by the
evolution of the spin system ($H_S+H_S^{(b)}+H_{SB}$) uncoupled
from the cantilever.

The functional derivative
${\delta\langle\sigma_Z(t)\rangle}/{\delta f(t_1)}$ in
Eq.~(\ref{dir}) is the response of the spin system to the weak
low-frequency external force $f(t)$. It has the magnetic
susceptibility $\chi_{zz}(\omega)$ of the spin system as its
Fourier transform:
\begin{equation}\label{chi}
\frac{\delta\langle\sigma_Z(t)\rangle}{\delta
f(t_1)}=\int\frac{d\omega}{2\pi}\exp[{-i\omega(t-t_1)}]\chi_{zz}(\omega)\;
,
\end{equation}
From Eqs.~(\ref{dir}) and (\ref{inter}) we obtain the equation of
motion of the cantilever:
\begin{multline}\label{cant1}
    \ddot z + \gamma _c \dot z + \omega _c^2 z =
     \lambda\langle\sigma_Z^{(0)}(t)\rangle
     -\frac{4\lambda}{d}\langle\sigma_Z^{(0)}(t)\rangle z\\+\lambda^2m\int
    dt_1\frac{\delta\langle\sigma_Z(t)\rangle}{\delta f(t_1)}
    z(t_1)
    +f_0\cos\nu t+{\eta}(t)\; ,
\end{multline}
Let us now analyze this equation in detail. The first two terms in
the right-hand side of Eq.~(\ref{cant1}) contribute, respectively,
to the amplitude and the frequency shifts due to the spins. The
modulation of either of these terms is usually employed in MRFM
experiments. In particular, the second term of the right-hand side
of Eq.~(\ref{cant1}) can be easily converted to the frequently
used expression for the frequency shift \cite{Mam}:
$$
\Delta f=m_z f_c (d^2B_z/dz^2)/2k_c\; ,
$$
where $m_z$ is the magnetic moment of the sample. These two terms
describe the direct influence of the spin on cantilever motion.

The third term in the right-hand side of Eq.~\!(\ref{cant1})
describes the \emph{additional back-action of the spin on the
cantilever motion}, which is due to the modification of the spin
dynamics by the cantilever. This term, which is the main subject
of our study here, gives rise to an additional frequency shift and
an additional damping of the cantilever.

Assuming that the steady-state value of the quantity
$\langle\sigma_Z^{(0)}(t)\rangle$ is independent of time, we
convert Eq.~\!(\ref{cant1}) to Fourier components of $z(t)$ [using
$z(t)=\int e^{-i\omega t}z(\omega) d\omega$)]:
\begin{multline}\label{zomeg}
  \Big\{\omega _c^2  - \omega ^2+\frac{4\lambda}{d}\langle\sigma_Z^{(0)}\rangle
-\lambda^2 m\chi'_{zz}(\omega)\Big.\\ - \Big.i\left[\omega \gamma
_c +\lambda^2 m\chi''_{zz}(\omega)\right]\Big\}z(\omega ) =
\frac{f_0}{2}\delta(\omega-\nu)+\eta(\omega)\; ,
\end{multline}
where we introduce the real and imaginary parts of the spin
susceptibility
$$
\chi_{zz}(\omega)=\chi_{zz}'(\omega)+i\chi_{zz}''(\omega)\; .
$$

Analyzing the third and the fourth terms of the rhs of
Eq.~\!(\ref{zomeg}) we see that the influence of the spins on the
cantilever produces the frequency shift
\begin{equation}\label{freqshift}
    \Delta\omega=\frac{2\lambda}{d\omega_c}\langle\sigma_Z^{(0)}\rangle
    -\frac{\lambda^2 m\chi'_{zz}(\omega)}{2\omega_c}\; ,
\end{equation}
where the first term in the rhs of Eq.~\!(\ref{freqshift})
represents the \emph{direct} contribution of the spins to the
frequency shift, while the second term in the rhs of
Eq.~(\ref{freqshift}) is the additional contribution to the
frequency shift that results from the \emph{indirect} influence of
the back-action of the spins on the cantilever motion.

From the fifth term in the rhs of Eq.~(\ref{zomeg}),
$-i\left[\omega \gamma _c +\lambda^2
m\chi''_{zz}(\omega)\right]z(\omega )$, we can write the total
damping of the cantilever:
\begin{equation}\label{gtot}
\gamma_{\rm total}=\gamma_c+\gamma_{\rm spin}\; ,
\end{equation}
where $\gamma_{\rm spin}$ is the frequency-dependent contribution
of the back-action of the spins to the damping of the cantilever:
\begin{equation}\label{gspin}
\gamma_{spin}=\frac{\lambda^2 m\chi''_{zz}(\omega)}{\omega}\; .
\end{equation}
From this expression we obtain a spin back-action-induced
modification of the cantilever quality factor:
\begin{equation}\label{qfac}
    \frac{1}{Q}=\frac{1}{Q_c}+\frac{\lambda^2
    m\chi''_{zz}(\omega)}{\omega_c^2}\; .
\end{equation}

\section{The low-frequency magnetic susceptibility of irradiated spins}
\label{lfsusc} The interaction of a two level system with an
external electromagnetic field tuned near the resonance of this
two-level system can be described using the dressed-state
approach\cite{Tan}, which recently was successfully applied to
investigate the interaction between solid state superconducting
qubits and an external radiation source\cite{Li,Green1}. Some
results obtained in Ref. \!\onlinecite{Green1} within the
dressed-state approach will be applied here for the investigation
of the interaction of a MRFM cantilever with irradiated nuclear
spins.

The energy levels of  a spin-1/2 interacting with a high-frequency
$\omega_{\rm mw}$ field has a term proportional\cite{Tan} to the
number $N_{\rm ph}$ of photons, with an additional splitting of
each photon state by the Rabi energy $\hbar\Omega_R/2$:
\begin{equation}\label{level}
    E^{\pm}(N_{\rm ph})=\hbar \omega_{\rm mw}N_{\rm ph}\pm \frac{1}{2}\hbar
    \Omega_R\; ,
\end{equation}
where $\Omega_R$ is the Rabi frequency

\begin{equation}\label{RF}
   \Omega_R=\sqrt{\delta^2+\Omega_1^2}\; ,
\end{equation}
with
$$
\Omega_1=\omega_1\langle N_{\rm ph}\rangle^{1/2};
$$
where $\langle N_{\rm ph} \rangle$ is the average number of high
frequency photons\cite{Comm1}, and
$$
\delta=\omega_{\rm mw}-\omega_0
$$
is the high-frequency detuning. For definitiveness, here we assume
$\delta>0$. The frequency $\Omega_1$ is directly related to the
amplitude of microwave field: $\Omega_1=\gamma B_1$.

As was shown in Ref.~\onlinecite{Green1}, this system can be
described, within the RWA approach, by the rate equations for the
elements of the reduced density matrix, which describe the
transition between Rabi levels [the levels $E^{\pm}(N_{\rm ph})$
in Eq.~(\ref{level}), with the same $N_{\rm ph}$].

\begin{equation}\label{ro}
    \frac{d\rho}{dt}=-A_1\rho+B\rho_++(\Gamma_-)\cos2\theta\; ,
\end{equation}

\begin{equation}\label{ro+}
    \frac{d\rho_+}{dt}=-i\Omega_R\rho_-+B\rho-A_2\rho_++(\Gamma_-)\sin2\theta\; ,
\end{equation}

\begin{equation}\label{ro-}
     \frac{d\rho_-}{dt}=-i\Omega_R\rho_+-\Gamma_{\varphi}\rho_-\; ,
\end{equation}

where
\begin{equation}\label{A1}
    A_1=\left[\frac{1}{T_1}\cos^22\theta+\Gamma_{\varphi}\sin^22\theta\right]\; ,
\end{equation}
\begin{equation}\label{A2}
    A_2=\left[\frac{1}{T_1}\sin^22\theta+\Gamma_{\varphi}\cos^22\theta\right]\;
    ,
\end{equation}
\begin{equation}\label{B}
    B=\left[\Gamma_{\varphi}-\frac{1}{T_1}\right]\sin2\theta\cos2\theta\;
    ,
\end{equation}
where $\Gamma_{\varphi}$ is the dephasing rate of a spin, which
can be expressed in terms of the spin-spin relaxation time $T_2$,
$\Gamma_{\varphi}=1/T_2$. Here, $T_1$ is the spin-lattice
relaxation time, which is related to up $\Gamma_\uparrow$ and down
$\Gamma_\downarrow$ transition rates between spin levels
$$
T_1^{-1}=\Gamma_\uparrow+\Gamma_\downarrow\ ;\ \ \
\Gamma_-=\Gamma_\uparrow-\Gamma_\downarrow\ .
$$

The angle $\theta$ is defined by $\tan2\theta=\!-\Omega_1/\delta$,
where $0<2\theta<\pi$, so that
$$
\cos2\theta=-\delta/\Omega_R\; ,
$$
and
$$
\cos\theta=\frac{1}{\sqrt{2}}\!\left(1-\frac{\delta}{\Omega_R}\right)^{1/2};\
\sin\theta=\frac{1}{\sqrt{2}}\!\left(1+\frac{\delta}{\Omega_R}\right)^{1/2}
$$

For equilibrium conditions, the relaxation $\Gamma_\downarrow$ and
excitation $\Gamma_\uparrow$ rates are related by the detailed
balance law:
\begin{equation}\label{balance}
\Gamma_{\uparrow}=\Gamma_{\downarrow}\exp\left(-\frac{\hbar\omega_0}{k_BT}\right)\;
.
\end{equation}
From Eq.~(\ref{balance}) we obtain
\begin{equation}\label{balance1}
   T_1\Gamma_-=-\tanh\left(\frac{\hbar\omega_0}{2k_BT}\right)\; .
\end{equation}
The quantity $\rho$ in Eq.~(\ref{ro}) is defined as the difference
of the populations between the higher and the lower Rabi levels.

The steady-state solution
($\frac{d\rho}{dt}=\frac{d\rho_-}{dt}=\frac{d\rho_+}{dt}=0$) for
Eqs.~(\ref{ro}), (\ref{ro+}), and (\ref{ro-}) is as follows:
\begin{equation}\label{st0}
    \rho^{(0)}=\frac{\left(\Gamma_{\varphi}^2+\Omega_R^2\right)}
    {\frac{\Gamma_{\varphi}^2}{T_1}+A_1\Omega_R^2}(\Gamma_-)\cos2\theta\;
    ,
\end{equation}

\begin{equation}\label{st+}
    \rho_+^{(0)}=\frac{\Gamma_{\varphi}^2}{\frac{\Gamma_{\varphi}^2}{T_1}+A_1\Omega_R^2}
    (\Gamma_-)\sin2\theta\; ,
\end{equation}

\begin{equation}\label{st-}
    \rho_-^{(0)}=-i\frac{\Omega_R}{\Gamma_{\varphi}}\rho_+^{(0)}\;
    .
\end{equation}

It is interesting to note that under high-frequency irradiation,
the population of the Rabi levels becomes inverted. This is seen
from Eq.~(\ref{st0}), where the quantity $\rho^{(0)}$ is positive,
since for $\delta>0$ we have $\cos2\theta=-\delta/\Omega_R<0$, and
always $\Gamma_-<0$.

In addition, as $\delta$ tends to zero, $\rho^{(0)}\rightarrow 0$,
which causes the equalization of the population of the two levels
when the high frequency irradiation is in exact resonance with the
NMR frequency $\gamma B_0$.

The quantity $\langle\sigma_Z(t)\rangle$ which is, by definition,
the longitudinal magnetization of a sample with $N$ spin-$1/2$
particles (see Appendix A) can be expressed\cite{Green1} in terms
of the matrix elements $\rho$ and $\rho_+$ :
\begin{equation}\label{sigma-z}
     \langle\sigma_Z(t)\rangle=
     \rho(t)\cos2\theta+\rho_+(t)\sin2\theta\; .
\end{equation}

Therefore, from the definition of $\chi_{zz}(\omega)$ in
Eq.~(\ref{chi}) we obtain
\begin{equation}\label{chizz11}
    \chi_{zz}(\omega)=\chi_{\rho}(\omega)\cos2\theta+\chi_{\rho_+}(\omega)\sin2\theta\;
    ,
\end{equation}
where $\chi_{\rho}(\omega)$ and $\chi_{\rho_+}(\omega)$ are the
spectral components of the response of $\rho(t)$ and $\rho_+(t)$
to a weak external force, and are defined similarly to
Eq.~(\ref{chi}).

The susceptibilities $\chi_{\rho}(\omega)$ and
$\chi_{\rho_+}(\omega)$ can be readily  found by investigating the
response of the reduced density matrix in Eqs.~(\ref{ro}),
(\ref{ro+}), and (\ref{ro-}) to a weak external perturbation [see
Eqs.~\!(\ref{hiro}), (\ref{hiro+}), and (\ref{hiro-}) in Appendix
B]. Hence, the expression for $\chi_{zz}(\omega)$ becomes:
\begin{equation}\label{chizz}
    \chi_{zz}(\omega)=\frac{\delta\,\Omega_1\,\Omega_R}{\hbar\,
    D(\omega)\,\Gamma_{\varphi}^2}\,
   \rho_+^{(0)}\,(-i\omega+2\Gamma_{\varphi})\; ,
\end{equation}
where $D(\omega)$ is given in Eq.~(\ref{Domega}).

The quantity $\rho_+^{(0)}$ in Eq.~(\ref{chizz}) is the
steady-state value (\ref{st+}) which can be written as follows:
\begin{equation}\label{st+1}
    \rho_+^{(0)}=-\frac{\Omega_1}{\Omega_R}\frac{1}{1+(\delta T_2)^2+T_2
    T_1\Omega_1^2}\tanh\left(\frac{\hbar\omega_0}{2k_BT}\right)\;
    ,
\end{equation}

and can be expressed in terms of the stationary magnetization
(\ref{mzst}):
\begin{equation}\label{mzst1}
       \frac{N\hbar\gamma}{2}\rho_+^{(0)}=-\frac{M_Z^{(st)}}{1+(\delta
       T_2)^2}\frac{\Omega_1}{\Omega_R}\; .
\end{equation}

\subsection{Resonance of the longitudinal magnetization at the Rabi frequency}

Historically, the detection of NMR is based on the Faraday law of
induction\cite{Pac}. That is why most of the measurement schemes
in NMR are based on the detection of the \emph{transverse}
magnetization, which oscillates with a relatively high frequency.

The detection of the \emph{longitudinal} magnetization is less
common because it requires measurements in a low frequency range
with a low signal-to-noise ratio. However, this drawback can be
circumvented by some techniques, such as the pre-polarization of a
sample in high field\cite{App}, or the use of superconducting
quantum interference devices (SQUIDs)\cite{Green3}, which allow to
obtain, in the micro- and nano-Tesla range, a resolution which is
beyond what is usually achieved in conventional high-field
NMR\cite{Mac,Mac1,Zot,Zot1,Burg}.
\begin{figure}
  \includegraphics[width=8 cm]{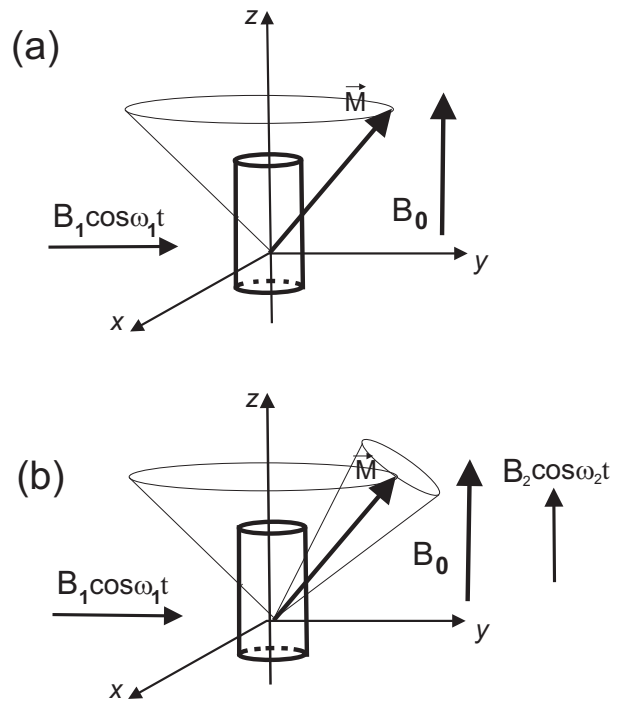}\\
  \caption{Resonance of the longitudinal magnetization $M_z$ at the Rabi frequency
   $\Omega_R$.
  (a) For conventional NMR, the longitudinal magnetization
   of a sample polarized in the field $B_0$ and
  subject to a circularly polarized excitation field $B_1\cos\omega_1
  t$, where $\omega_1\approx\omega_0$, exhibits a precession
  around the $z$-axis with a time-independent steady-state
  $z$-projection of the longitudinal magnetization $M_z$.
  (b) A second low-frequency excitation $B_2\cos\omega_2 t$, where $\omega_2\approx\Omega_R$, applied
   along the $z$-axis, produces resonant oscillations of $M_z$ near
   the Rabi frequency $\Omega_R$.
   }\label{resrabi}
\end{figure}

In MRFM there is no choice other than to measure the longitudinal
component of the nuclear polarization, since the resonance
frequencies of MRFM cantilevers are well below those corresponding
to the frequencies of NMR transitions. From this point of view, it
is interesting to note that the \emph{longitudinal} magnetization
$M_z$ of a sample placed in a high frequency resonant radiation
field \emph{shows a clear resonance at the Rabi frequency if the
sample is subject to an additional low-frequency excitation}
directed along the $z$-axis with energy (see Fig.~\ref{resrabi})
$$
H_{\rm LF}=\hbar\gamma B_2\cos\omega t\ .
$$

As shown in Refs.~\onlinecite{Green1} and \onlinecite{Green4},
this effect is a general feature of any two-level dissipative
system. In this case, the low-frequency evolution of the
longitudinal magnetization can be expressed in terms of the spin
susceptibility $\chi_{zz}(\omega)$ (\ref{chizz}):
\begin{equation}\label{mz2}
    M_Z(t)=M_Z^{(\rm st)}+\frac{N\hbar\gamma}{2}\hbar\gamma
    B_2\Big[{\chi}'_{zz}(\omega)\cos\omega t
    +{\chi}''_{zz}(\omega)\sin\omega t\Big],
\end{equation}
where $M_Z^{(\rm st)}$ is given by Eq.~\!(\ref{mzst}).

With the aid of Eq.~\!(\ref{mzst1}) we rewrite Eq.~\!(\ref{mz2})
in the following form:
\begin{equation}\label{mz1}
    M_Z(t)=M_Z^{(\rm st)}\left\{1+\frac{B_2}{B_1}\left[\widetilde{\chi}'_{zz}(\omega)\cos\omega
    t+\widetilde{\chi}''_{zz}(\omega)\sin\omega t\right]\right\},
\end{equation}
where
\begin{equation}\label{chiz}
    \widetilde{\chi}_{zz}(\omega)=-\frac{\delta\,\Omega_1^3
    (-i\omega+2\Gamma_{\varphi})}{D(\omega)\,\Gamma_{\varphi}^2\,
    (1+(T_2\delta)^2)}\text{  ,}
\end{equation}
$$
T_2=1/\Gamma_{\varphi},\ \ \ \ \Omega_1=\gamma B_1\,,
$$
$B_1$ is the amplitude of the \emph{high}-frequency resonance
excitation, and $B_2$ is the amplitude of the \emph{low}-frequency
signal which excites the Rabi oscillations of the longitudinal
magnetization $M_z$.

\begin{figure}
  \includegraphics[width=\columnwidth]{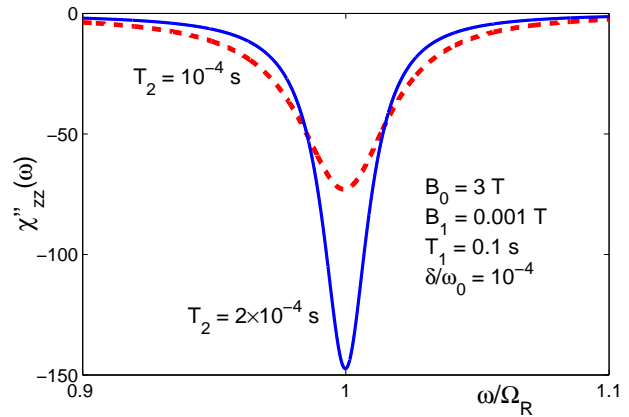}\\
  \caption{(Color online) The dependence of dissipative part
  $\widetilde{\chi}''_{zz}(\omega)$ of spin susceptibility
   for low frequencies $\omega\approx\Omega_R$.}\label{rabires}
\end{figure}

The resonance at the Rabi frequency $\Omega_R$ is clearly seen in
Fig.~\ref{rabires}, where we plot (for two values of the spin-spin
relaxation time $T_2$) the dissipative part of the spin
susceptibility (\ref{chiz}) as a function of the low-frequency
$\omega$ ($\omega\approx\Omega_R$).

It should be remembered that our linear approximation is valid
within the range
\begin{equation}\label{valrange}
    \frac{N\hbar\gamma}{2}\hbar\gamma B_2\,|{\chi_{zz}(\omega)}|\,<M_Z^{(\rm
    st)}\; ,
\end{equation}
from where we obtain the range of the amplitudes of the
low-frequency signal $B_2$, where the expression (\ref{mz1}) is
consistent with the condition (\ref{valrange}):
$$
B_2<B_1/\widetilde{\chi}''_{zz,\,\rm max}(\omega)
$$

Hence, \emph{near the Rabi resonance (i.e.,
$\omega\approx\Omega_R$), a significant modulation of the
longitudinal magnetization $M_z$ can be induced by a low frequency
drive. }

The resonance of the longitudinal magnetization described above
is, in some sense, a \emph{low-frequency analog of conventional
NMR}, where the resonance of the transverse nuclear magnetization
is being measured at a high resonance frequency.

\section{Influence of spins on the frequency shift and the damping of the cantilever}
\label{influence}
\subsection{Influence of the driven equilibrium longitudinal magnetization on the
damping and frequency shift of the cantilever.}

Here we  consider the situation where, before the measurements
start, the spins under the application of a polarizing external
magnetic field reach thermal equilibrium with their environment.
In other words, in this case, the spin-lattice relaxation time
$T_1$ is sufficiently short (for example, in the millisecond
range) to ensure the application of conventional NMR measurement
protocols. For this case, the corresponding susceptibilities are
given by the expressions (\ref{hiro1}), (\ref{hiro+1}), and
(\ref{hiro-1}).

In order to analyze  the damping and the frequency shift of the
cantilever in Eq.~(\ref{zomeg}), we explicitly write down the real
and imaginary parts of $\chi_{zz}(\omega)$ from Eq.~(\ref{chizz}):
\begin{equation}\label{rechizz}
\chi_{zz}'(\omega)=\frac{\delta\Omega_1\Omega_R}{\hbar\Gamma_{\varphi}^2}\rho_+^{(0)}
\frac{2\Gamma_{\varphi}D_1(\omega)-\omega
D_2(\omega)}{D_1^2(\omega)+ D_2^2(\omega)}\; ,
\end{equation}
\begin{equation}\label{imchizz}
\chi_{zz}''(\omega)=-\frac{\delta\Omega_1\Omega_R}{\hbar\Gamma_{\varphi}^2}\rho_+^{(0)}
\frac{\omega
D_1(\omega)+2\Gamma_{\varphi}D_2(\omega)}{D_1^2(\omega)+
D_2^2(\omega)}\; ,
\end{equation}
where
\begin{equation}\label{d1}
    D_1(\omega)\equiv {\rm Re}[D(\omega)]=\frac{1}{T_1}[A_1
    T_1\Omega_R^2+\Gamma_{\varphi}^2-\omega^2(1+2T_1\Gamma_{\varphi})]\;
    ,
\end{equation}
\begin{equation}\label{d2}
    D_2(\omega)\equiv
    {\rm Im}[D(\omega)]=\omega\left(\omega^2-\Omega_R^2-\Gamma_{\varphi}^2-
    \frac{2\Gamma_{\varphi}}{T_1}\right).
\end{equation}
It is worth noting that at the exact resonance ($\delta=0$), the
back-action influence of the spins on the cantilever vanishes,
since the corresponding susceptibilities (\ref{rechizz})  and
(\ref{imchizz}) are equal to zero at this point. This is a
consequence of the equalization of the population of Rabi levels
at the point of exact resonance ($\rho^{(0)}$ tends to zero as
$\delta$ approaches zero). This produces the \emph{vanishing of
the energy flow between the spins and the cantilever at the Rabi
frequency.}

Another point is that the sign of the susceptibility
$\chi''_{zz}(\omega)$ in Eq.~\!(\ref{imchizz}) is \emph{opposite}
to that of $\delta$, for any value of its parameters. This follows
from the fact that the numerator $\omega
D_1(\omega)+2\Gamma_{\varphi}D_2(\omega)$ in Eq.~(\ref{imchizz})
is always negative, and the quantity $\rho_+^{(0)}$ is also
negative. Therefore, for $\delta>0$ (when the higher Rabi level is
more populated than the lower) the contribution of the spins to
the damping of the cantilever is negative [see Eq.~(\ref{zomeg})].
In this case, \emph{the cantilever is being heated by absorbing
the Rabi photons emitted by the spins.}

In the opposite case, when $\delta<0$ (i.e., the higher Rabi level
is less populated than the lower) the contribution of the spins to
the damping of the cantilever is positive, therefore \emph{cooling
the cantilever which gives up Rabi photons to the spin system.}

It is worthwhile to consider the dependence of the dissipative
part of susceptibility $\chi''_{zz}(\omega)$
(Eq.~\!(\ref{imchizz})) on the frequency $\omega$. It has two
peaks. A lorentzian peak is in the vicinity of the Rabi frequency
$\Omega_R$, as it is evident from (\ref{d2}). The condition for
this is the relative large high-frequency detuning
($\delta\gg\Gamma_{\varphi}, \Omega_1$). The approximate
expression for the peak value at the Rabi resonance is as follows:
\begin{equation}\label{peak}
\chi_{zz}''(\omega)\Big|_{peak}\approx\rho_+^{(0)}\frac{\delta\Omega_1}{2\hbar
\Gamma_{\varphi}^3}
\end{equation}

The other peak is related to the spin-lattice relaxation time
$T_1$ of the longitudinal magnetization. It lies at much lower
frequencies ($\omega\ll\Omega_R$). If we assume
$\omega\ll\Omega_R, \Gamma_{\varphi}$; $\delta\gg\Gamma_{\varphi},
T_1, \Omega_1$; $\Gamma_{\varphi}T_1\gg1$, we obtain
$D_1(\omega)\approx\delta^2/T_1$,
$D_2(\omega)\approx-\delta^2\omega$, which yields the following
expressions for the susceptibility:
\begin{equation}\label{rechizzlowfrequ}
\chi_{zz}'(\omega)\approx\rho_+^{(0)}\frac{\delta}{|\delta|}\frac{2\Omega_1T_1}{\hbar\Gamma_{\varphi}}\frac{1}{1+\omega^2T_1^2}
\end{equation}
\begin{equation}\label{imchizzlowfrequ}
\chi_{zz}''(\omega)\approx\rho_+^{(0)}\frac{\delta}{|\delta|}\frac{2\Omega_1T_1}{\hbar\Gamma_{\varphi}}\frac{\omega
T_1}{1+\omega^2T_1^2}
\end{equation}
with the maximum of $\chi_{zz}''(\omega)$ being at
$\omega_{\texttt{max}}=1/T_1$. These expressions are analogous to
the Debye formulae for the low-frequency dispersion of the
dielectric constant. In the context of NMR, this Debye-like
behavior is known for the response of the transverse magnetization
in a weak polarizing field\cite{Abr}.

To estimate the magnitude of this effect, we take as a guide the
experimentally accessible parameters from Ref.~\onlinecite{Mam}.
From Fig.~2b of that paper we take the following values of the
magnetic field $B_F{(0)}$ on the tip, $B_F{(0)}=100$ mT, and a
corresponding value of $d$, the distance between the tip and the
sample: $d=60$ nm. From the spring constant $k_c=6\times 10^{-5}$
N/m and the resonance frequency of the cantilever, $f_c=3$ kHz, we
estimate the cantilever mass
$$
m=k_c/\omega_c^2=1.6\times 10^{-13}\,\rm kg.
$$
For the element $^{19}$F, which was the subject of study in
Ref.~\onlinecite{Mam}, the nuclear gyromagnetic ratio
$\gamma=2\pi\times 40$ MHz/T. This allows to estimate the coupling
factor $\lambda$:
$$
\lambda=\frac{3N\hbar\gamma B_F(0)}{2md}\approx \Big(4.12\times
10^{-7}\times N\Big)\, \frac{\rm m}{\rm s^2}\, .
$$
Here $N$ is the number of spins in the resonant slice. Thus, for
the factor $\lambda^2 m/\hbar$ we obtain the estimate
$$
\frac{\lambda^2 m}{\hbar}\approx \Big(2.6\times 10^8\times
N^2\Big)\,\rm s^{-3}\; .
$$
In addition, we take $T_1=10^{-3}$ s,
$T_2=1/\Gamma_{\varphi}=10^{-6}$ s, $T=0.6$ K, $B_0=3$ T,
$B_1=\Omega_1/\gamma=10^{-3}$ T.

\begin{figure}
  \includegraphics [width=\columnwidth]{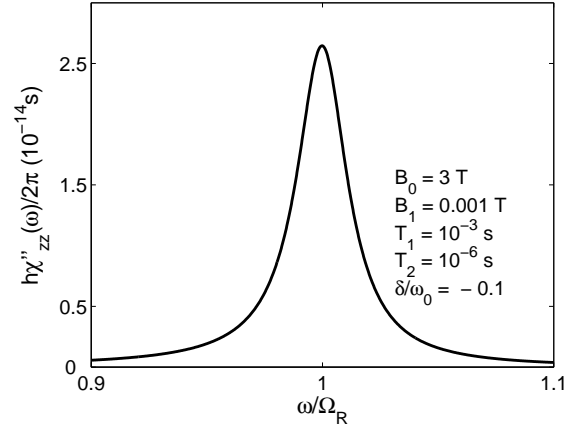}\\
  \caption{The imaginary part of spin susceptibility as function of normalized frequency $\omega/\Omega_R$
  in the vicinity of Rabi frequency, for $T=0.6$ K, N=10$^5$.}\label{imsus1}
\end{figure}

\begin{figure}
  \includegraphics [width=\columnwidth]{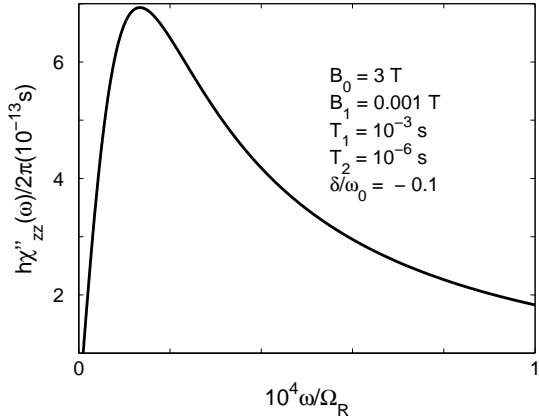}\\
  \caption{The imaginary part of spin susceptibility as a function of the normalized frequency
  $\omega/\Omega_R$,
  in the vicinity of the Debye-like peak, for $T=0.6$ K, and N=10$^5$.}\label{imsus1a}
\end{figure}

The contribution of the spins to the cantilever \emph{damping}
$\gamma_{\rm spin}$ is proportional to the imaginary part of the
spin system susceptibility $\chi''_{zz}(\omega)$ [see Eq.
(\ref{gspin})]. The susceptibility $\chi''_{zz}(\omega)$ has a
clear resonance at the Rabi frequencies, which is shown in
Fig.~\ref{imsus1}. The second Debye-like peak is shown in
Fig.~\ref{imsus1a}.
It is worthwhile to note that for the parameters we used here the
peak value in Fig.~\ref{imsus1a} is much higher than the Rabi peak
in Fig.~\ref{imsus1}.

\begin{figure}
 \includegraphics[width=\columnwidth]{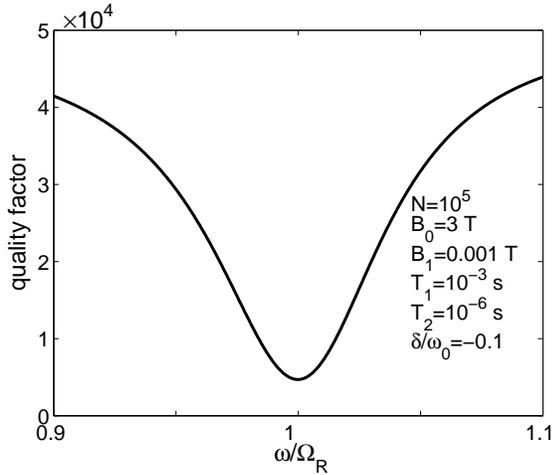}\\
  \caption{The modified quality factor $Q$ of the cantilever near the Rabi frequency $\Omega_R$. Here, the quality factor of
   the unloaded cantilever is $Q_c=5\times 10^4$, $T$=0.6 K.}\label{Qual}
\end{figure}

\begin{figure}
 \includegraphics[width=\columnwidth]{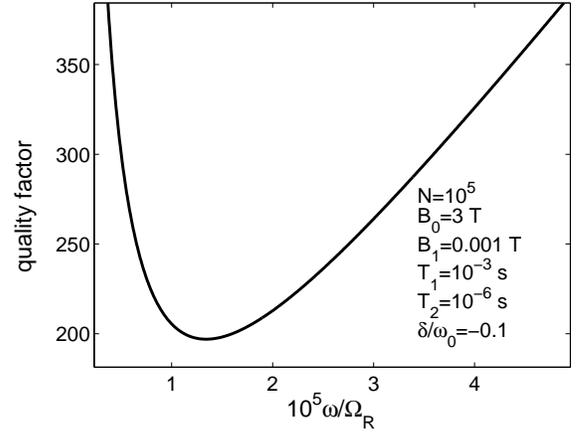}\\
  \caption{The modified quality factor $Q$ of the cantilever near the Debye-like peak. Here, the quality factor $Q_c$ of
   the unloaded cantilever is $Q_c=5\times 10^4$, and $T$=0.6 K.}\label{Qual1}
\end{figure}

These resonances modify within the corresponding frequency range
the bare cantilever quality factor $Q_c$ in Eq.~(\ref{qfac}), as
shown in Fig.~\ref{Qual} and Fig.~\ref{Qual1}.


\begin{figure}
  \includegraphics[width=\columnwidth]{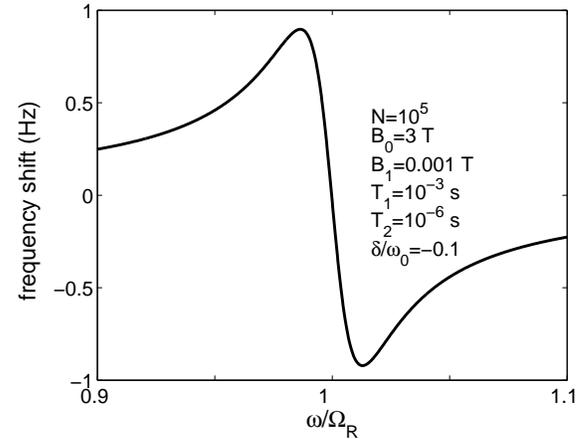}\\
  \caption{Frequency shift,
  $\lambda^2m\chi'_{zz}(\omega)/2\omega_c$, due to the back-action
  of the spins,
  versus the normalized frequency
  $\omega/\Omega_R$, for $T=0.6$ K.}\label{frshift}
\end{figure}

\begin{figure}
  \includegraphics[width=\columnwidth]{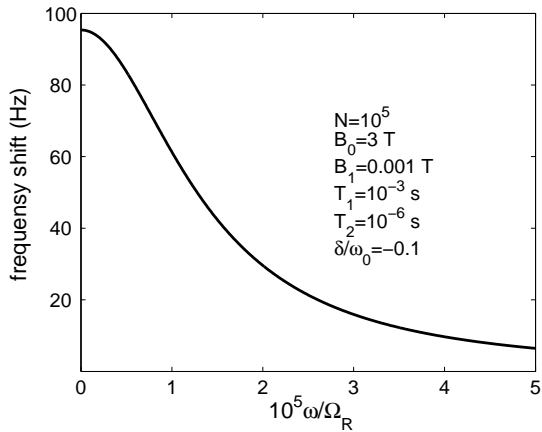}\\
  \caption{Frequency shift,
  $\lambda^2m\chi'_{zz}(\omega)/2\omega_c$, near the Debye-like peak, for $T=0.6$ K.}\label{frshiftlow}
\end{figure}

The contribution to the cantilever \emph{frequency shift} is given
by the real part of the spin system susceptibility
$\chi'_{zz}(\omega)$ in Eq.~(\ref{rechizz}). The associated
frequency shift, $\lambda^2m\chi'_{zz}(\omega)/2\omega_c$, near
the Rabi resonance is shown in Fig.~\ref{frshift}, and near the
Debye-like peak in Fig.~\ref{frshiftlow}


For the parameters used here, the Rabi frequency $\Omega_R\approx
12$ MHz and the Debye-like peak in Fig.~\ref{imsus1a} is located
near 160 Hz. Between these two peaks, the quality factor is
gradually increased from about $200$ to almost its bare value of
$50,000$ before it falls down to $5,000$ at the Rabi frequency.
The resonance frequency of the cantilever ($f_c=3$ KHz) lies at
the continuation of the right side of the curve shown in
Fig.~\ref{Qual1}. The calculations show that at $f=3$ KHz the
quality factor $Q\approx1,800$ with the frequency shift $\Delta
f\approx 0.32$ Hz, which is in the range of the modified bandwidth
($\approx 2$ Hz). In principle, the dissipative part of the
susceptibility $\chi''_{zz}(\omega)$ is very sensitive to its
parameters, especially, to $\delta$, $\Gamma_{\varphi}$, and
$\Omega_1$. If we had taken, for example,
$T_2=1/\Gamma_{\varphi}=10^{-7}$ s, other parameters being
unchanged, we would obtain $Q\approx 200$ near $3$ kHz, the
resonance of the cantilever. Hence, while in the real experiment
the external parameters, such as $\delta$ and $\Omega_1$ can be
controlled, the estimation of the damping effect at the given
frequency requires the knowledge, with good accuracy, of the
spin-spin relaxation time $T_2$ .

\subsection{Effect of the nuclear spin noise on the damping of the cantilever}

If the spin-lattice relaxation time $T_1$ is extremely long, which
happens at low temperatures, it is not possible to use
conventional NMR methods, which rely on the measurement of the
equilibrium spin polarization. In addition, for nanoscale volumes
(below about (100nm)$^3$) the statistical spin polarization
exceeds the mean Boltzmann polarization\cite{Deg1}. Hence, an
alternative approach would be to measure a naturally-occurring
statistical polarization: the spin noise\cite{Mam1, Mam, Deg}. In
this case, a sample has a magnetic moment with a
mean-squared-value proportional to $\sqrt{N}\mu$, where $N$ is the
number of nuclear spins in the resonant slice, and $\mu$ is the
magnetic moment of a single particle\cite{Bloch, Deg1}.

Hence, for long $T_1$ we take for $\rho(t')$, $\rho_+(t')$, and
$\rho_-(t')$, in Eqs. (\ref{hiro}), (\ref{hiro+}), and
(\ref{hiro-}) their \emph{initial} values: $\rho{(0)}$,
$\rho_+{(0)}$ , and $\rho_-{(0)}$. These values correspond to
natural spin fluctuations in a sample which is in thermal
equilibrium and under no external influence ($\omega_0=0,
 \Omega_1=0$). In order to find $\rho{(0)}$,
$\rho_+{(0)}$, and $\rho_-{(0)}$, we put $\Omega_1=0$ in Eqs.
(\ref{st0}), (\ref{st+}), and (\ref{st-}). In this case the Rabi
levels disappear and we get a sample in a constant polarizing
field $B_0$ with $\rho^{(0)}=-T_1\Gamma_-$, the equilibrium
normalized population difference between two levels (see Eq.
(\ref{balance1})).  Hence, for this case, the quantity
$\rho^{(0)}$ which before was  the population difference between
Rabi levels, remains the population difference between levels of a
spin in the field $B_0$. Therefore, in the absence of an external
field ($B_0=0$) it is reasonable to consider $\rho^{(0)}$ as
$\rho{(0)}$, the apparent normalized population difference which
provides the mean-squared-value of the naturally-occurring
magnetic moment $\sqrt{N}\mu$. Hence,

$$
\rho{(0)}=-\frac{\sqrt{N}\mu}{N\mu }=-\frac{1}{\sqrt{N}}\
$$
Here, the quantity $\rho{(0)}$ is negative because the upper level
is now less populated than the lower one.

In the same limit ($\Omega_1=0$) we obtain from (\ref{st+}), and
(\ref{st-}) $\rho^{(0)}_+=\rho^{(0)}_-=0$. This result does not
depend on $B_0$ and remains unchanged when $B_0=0$. The quantities
$\rho^{(0)}_+$ and $\rho^{(0)}_-$ describe the density matrix
elements between Rabi levels (see Eqs. \!(26) in\cite{Green1})
which should tend to zero when Rabi levels disappear. Hence, in
the absence of the polarizing field $B_0$, it is reasonable to
consider $\rho^{(0)}_+=\rho_+{(0)}=0$ and
$\rho^{(0)}_-=\rho_-{(0)}=0$.

Therefore, for the corresponding susceptibilities, we obtain the
expressions (\ref{hiro2}), (\ref{hiro+2}), and (\ref{hiro-2}).
Hence, in this case the expression for $\chi_{zz}(\omega)$
becomes:
\begin{equation}\label{chizz1}
    \chi_{zz}(\omega)=\frac{\Omega_1^2}{\hbar\sqrt{N}\Omega_R d(\omega)}
    \left[\frac{\Gamma_{\varphi}\Omega_1^2}{\Omega_R^2}-i\left(\omega-\frac{\delta^2\Gamma_{\varphi}}{\Omega_R^2}\right)
    \right],
\end{equation}
where $d(\omega)$ is given in (\ref{dom}).

The use of the spin noise for the MRFM 2D and 3D image
reconstruction of the nuclear spin density was described in
Refs.~\onlinecite{Mam} and \onlinecite{Deg}. These papers briefly
reported that they measured an unexplained substantial decrease of
the quality factor of the cantilever (from 50,000 to 8,000 in
Ref.~\onlinecite{Mam}, and from 30,000 to several thousand in
Ref.~\onlinecite{Deg}). Here we show that, qualitatively,
\emph{this effect might be explained by the back-action of spin
noise on the cantilever motion.}

For the quantitative estimate of this effect, we take as a guide
the necessary parameters from Ref.~\onlinecite{Mam}. Here we now
assume  that the spin-lattice relaxation time $T_1$ is so long
that it prevents the manipulation of the equilibrium Boltzmann
polarization. For example, for the atom $^{19}$F in calcium
fluoride CaF$_2$ studied in Ref.~\onlinecite{Mam}, the time $T_1$
at the experimental \cite{Kuh} temperature $T=0.6$ K was about
10$^4$ s.

\begin{figure}
  \includegraphics[width=\columnwidth]{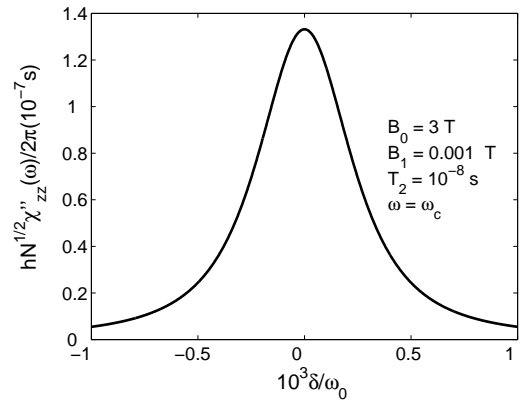}\\
  \caption{The dependence of the dissipative part of
the spin susceptibility on the high frequency detuning
   $\delta=\omega_{\rm mw}-\omega_0$, with $\omega_c/2\pi=3$ kHz.}\label{imsus3}
\end{figure}

The investigation of the imaginary part of the magnetic
susceptibility (\ref{chizz1}) shows that its dependence on the
high-frequency detuning $\delta$ shows a sharp peak at the point
of resonance, $\delta=0$ (see Fig.~\ref{imsus3}, where the
function $\chi_{zz}''(\omega)$ is drawn at the resonance of the
cantilever).
Thus, we now write down the real and imaginary parts of the
magnetic susceptibility (\ref{chizz1}) for the exact resonance
($\delta=0$), where we expect the maximum effect:
\begin{equation}\label{chizzre}
    \chi_{zz}'(\omega)=-\,\frac{\Omega_1}{\hbar\sqrt{N}
    |d(\omega)|^2}(\omega^2-\Omega_1^2)(\omega^2+\Gamma_{\varphi}^2)\;
    ,
\end{equation}
\begin{equation}\label{chizzim}
    \chi_{zz}''(\omega)=\frac{\omega\Omega_1\Gamma_{\varphi}}{\hbar\sqrt{N}
    |d(\omega)|^2}(\omega^2+\Gamma_{\varphi}^2)\; ,
\end{equation}
and here
\begin{equation}\label{d22}
    |d(\omega)|^2=\omega^2(\omega^2-\Omega_1^2-\Gamma_{\varphi}^2)^2+
    \Gamma_{\varphi}^2(\Omega_1^2-2\omega^2)^2\; .
\end{equation}
From Eq.~(\ref{chizzim}) it follows that the contribution of the
spins to the damping of the cantilever is always positive, which
means that the spin noise is cooling the cantilever.

\begin{figure}
  \includegraphics[width=\columnwidth]{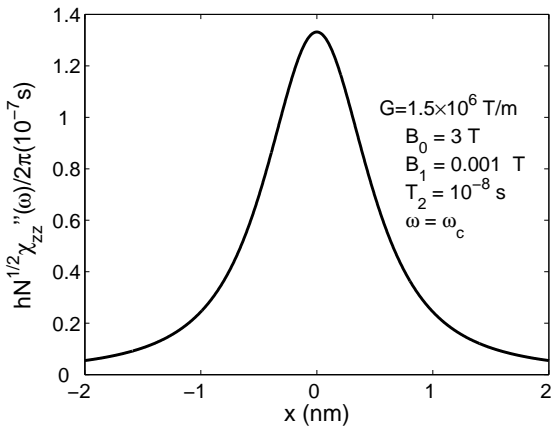}\\
  \caption{The dependence of the dissipative part of
the spin susceptibility on the distance $x$ from the point of
exact resonance, and here $\omega_c/2\pi=3$ kHz.}\label{slice}
\end{figure}

The width of the distribution $\Delta\delta$ shown in
Fig.~\ref{imsus3} is directly connected to the thickness of the
resonant slice $\Delta x=\Delta\delta/\gamma G$, where $G$ is the
magnetic field gradient. The dependence of the dissipative part of
the spin susceptibility on the distance from the point of exact
resonance is shown in Fig.~\ref{slice} for $B_0=3$ T and
$G=1.5\times 10^6$ T/m.
From this figure, we estimate the effective thickness of the
resonant slice as the full-width at half-maximum of the curve
shown in Fig.~\ref{slice}. Thus, we obtain $\Delta x\approx 1$ nm,
which corresponds to the width of high-frequency detuning
$\Delta\delta\approx 5\times 10^{-4}\omega_0$ (The full width at
half maximum of the curve in Fig.~\ref{imsus3}).

For example, if we take for the sample the dimensions 90
nm\,$\times$\,90 nm\,$\times$\,80 nm, with roughly 30 million
nuclear spins (see Ref.~\onlinecite{Mam}), we obtain the effective
number of spins in the resonant slice 90 nm\,$\times$\,90
nm\,$\times$\,1 nm that gives the main contribution to the effect:
$N_{\rm eff}\approx 3.5\times 10^5$.

Below we estimate, from Eq.~(\ref{chizzim}), the contribution of
the spin noise to the cantilever damping at the  resonance
frequency of the cantilever $\omega_c$, assuming
$\Gamma_{\varphi}\gg\omega/2\pi, \Omega_1/2\pi$. The result is
(see Eq.~\ref{qfac}):
\begin{equation}\label{qual}
    \frac{\lambda^2 m\chi''_{zz}(\omega)}{\omega_c^2}=\frac{\lambda^2
    m}{\sqrt{N}\hbar\omega_c^2\Gamma_{\varphi}}\frac{\Omega_1}{\omega_c} .
\end{equation}
In order to estimate the modified quality factor from
(\ref{qual}), we take $N\approx 10^5$, the number of spins in the
resonant slice, $\Gamma_{\varphi}\approx 10^8$ s$^{-1}$, $f_c=3$
kHz, and the bare $Q_c=5\times 10^4$. Hence, for zero
high-frequency detuning ($\delta=0$) we obtain from
Eq.~(\ref{qfac}) the modified quality factor $Q\approx 0.3$. This
enormous reduction of the quality factor is primarily due to the
large number of nuclei $N$ which are simultaneously at the exact
magnetic resonance ($\delta=0$).

However, we stress that, strictly speaking, our estimates above
cannot be considered as the only possible explanation of the
cantilever damping observed in Refs.~\onlinecite{Mam} and
\onlinecite{Deg}. One of the reasons for this is that the long
side of the cantilever in Refs.~\onlinecite{Mam} and
\onlinecite{Deg} was perpendicular to the sample surface, which is
different from the design shown in Fig.~\ref{cantilever}. For
their design, the damping of the cantilever due to the spin-noise
back-action is more sensitive (compared to our case) to the
density distribution of the spins over the sample surface. Another
reason is that here we assume that all resonant spins feel the
same field and are located at the same distance from the tip just
beneath it. Hence, to obtain more realistic values of $Q$, it is
necessary to modify Eq. (\ref{Bf}) with a more careful account of
the density distribution of the resonant spins over a sample.

\section{conclusion}
In this paper we investigate the interaction of the MRFM
cantilever with a system of nuclear spins. We show that the
back-action of nuclear spins results in an additional contribution
to the damping and frequency shift of the cantilever vibrations.
We also show that a spin system may significantly change the
quality factor of the cantilever. The cantilever can be either
heated or cooled, depending on the sign of the high-frequency
detuning. This effect exhibits a resonant nature, with a maximum
at the Rabi frequency. We show that the main reason for this
effect is that the longitudinal magnetization (which is commonly
measured in MRFM experiments) exhibits a resonance at the Rabi
frequency, which can significantly alter its low-frequency
evolution. We also analyze the influence of the spin noise on the
cantilever damping and show that the spin noise may lead to a
significant reduction of the quality factor of the cantilever.

The interesting question is the lower bound on the cooling of the
cantilever by nuclear spins.  It might seem at first that the
lower limit on the cooling of the cantilever is set by its
zero-point fluctuations. However, this is not the case. From
general considerations, the lower limit is set by the direct
contribution of the spin noise to the cantilever fluctuations
(see, for example, Ref.~\!\onlinecite{Graj}). In order to make a
reasonable estimate of this limiting temperature, it is necessary
to: a) first calculate the spectrum of the spin fluctuations under
a high-frequency field, and b) afterwards to consider a small
number of Rabi photons, which requires treating the cantilever
quantum mechanically. In our paper, we treat the spin system
quantum mechanically, while the number of Rabi photons is large,
which means that the cantilever behaves classically. This problem
will be the subject of future investigations.

\acknowledgements YaSG thanks A. Smirnov for valuable discussions
and acknowledges partial support from the Russian Foundation for
Basic Research, Grant RFBR-FRSFU No. 09-02-90419. EI acknowledges
the hospitality of RIKEN (Japan) and the financial support from
the EU through the EuroSQIP project, and from the Federal Agency
on Science and Innovations of Russian Federation under contract
No. 02.740.11.5067. FN acknowledges partial support from the
National Security Agency (NSA), Laboratory for Physical Sciences
(LPS), Army Research Office (ARO), National Science Foundation
(NSF) Grant No. EIA-0130383, and the JSPS-RFBR contract No.
06-02-91200.

\appendix
\section{Dynamics of an irradiated spin in the dressed state approach}

Here we very briefly summarize some results of
Ref.~\onlinecite{Green1}, as applied to the problem we study in
this paper.

In addition to Eqs.~(\ref{ro}), (\ref{ro+}), and (\ref{ro-})
(which describe the transition between Rabi levels of an
irradiated spin) the (spin+field) system can also be characterized
by the density matrix elements $\kappa$'s which describe the
transitions between levels whith photon numbers that differ by
one. These levels are approximately separated by $\hbar\omega_0$,
the energy between the levels of a bare spin. The rate equations
for the $\kappa$'s are \cite{Green1}:

\begin{equation}\label{k}
    \frac{d\kappa^+}{dt}=-i\omega_{mw}\kappa^+\; ,
\end{equation}

\begin{equation}\label{kap}
    \frac{d\kappa}{dt}=-i\omega_{mw}\kappa-A_1\kappa+B\kappa_++\kappa^+(\Gamma_-)\cos2\theta\;
    ,
\end{equation}

\begin{equation}\label{kap+}
    \frac{d\kappa_+}{dt}=-i\omega_{mw}\kappa_+-i\Omega_R\kappa_-+B\kappa-A_2\kappa_++\kappa^+(\Gamma_-)\sin2\theta\;
    ,
\end{equation}

\begin{equation}\label{kap-}
     \frac{d\kappa_-}{dt}=-i\omega_{mw}\kappa_--i\Omega_R\kappa_+-\Gamma_{\varphi}\kappa_-\;
     .
\end{equation}
For possible applications of this method, the quantities to be
measured are the averages of the Pauli spin operators
$\langle\sigma_X\rangle$, $\langle\sigma_Y\rangle$,
$\langle\sigma_Z\rangle$. As was shown in Ref.~\onlinecite{Green1}
\begin{equation}\label{sigma-z-a}
     \langle\sigma_Z\rangle=
     \rho(t)\cos2\theta+\rho_+(t)\sin2\theta\; ,
\end{equation}
\begin{equation}\label{sigma-x-a}
     \langle\sigma_X\rangle=
     \sin2\theta {\rm{Re}}[\kappa(t)]-\cos2\theta {\rm{Re}} [\kappa_+(t)]-{\rm{Re}}
     [\kappa_-(t)],
\end{equation}
\begin{equation}\label{sigma-y-a}
     \langle\sigma_Y\rangle=-\sin2\theta {\rm{Im}}[\kappa(t)]
     +\cos2\theta {\rm{Im}}[\kappa_+(t)]+{\rm{Im}}[\kappa_-(t)],
\end{equation}
These quantities are directly connected to the longitudinal
magnetization of a sample with $N$ spin-$1/2$ particles:
\begin{equation}\label{mz}
    M_Z=-\,\frac{N\gamma\hbar}{2}\langle\sigma_Z\rangle\; ,
\end{equation}

and its transverse components
\begin{equation}\label{mxy}
    M_X=-\frac{N\gamma\hbar}{2}\langle\sigma_X\rangle\ ;\ \ \
    M_Y=-\frac{N\gamma\hbar}{2}\langle\sigma_Y\rangle\ .
\end{equation}

It is very instructive here to show that the steady-state solution
for the density matrix provides the well-known Bloch expressions
for the longitudinal and transverse components of the
magnetization.

The stationary magnetization $M_Z^{(\rm st)}$, which is defined in
(\ref{mz}), is obtained from (\ref{sigma-z}) with the help of
(\ref{st0}), (\ref{st+}). By using the substitutions
$\Gamma_{\varphi}=1/T_2$, $\Omega_1=\gamma B_1$, we obtain for
$M_Z^{(\rm st)}$:

\begin{equation}\label{mzst}
    M_Z^{(\rm st)}=M_0\frac{1+\left(T_2\delta\right)^2}{1+\left(T_2\delta\right)^2+T_1T_2\left(\gamma
    B_1\right)^2}\; ,
\end{equation}
where
\[
M_0=\frac{N\hbar\gamma}{2}\tanh\left(\frac{\hbar\gamma
B_0}{2k_BT}\right)\ .
\]
Now we find the steady-state solutions of Eqs.~(\ref{k}),
(\ref{kap}), (\ref{kap+}), and (\ref{kap-}). It is not difficult
to see that the solution of these equations has the form:
$\kappa^+=c^+e^{-i\omega_{mw} t}$, $\kappa=ce^{-i\omega_{mw} t}$,
$\kappa_+=c_+e^{-i\omega_{mw} t}$, $\kappa_-=c_-e^{-i\omega_{mw}
t}$, where
\begin{equation}\label{c}
    c=c^+\frac{\Gamma_-}{D}\left(\Gamma_{\varphi}^2+\Omega_R^2\right)\cos2\theta\;
    ,
\end{equation}

\begin{equation}\label{c+}
     c_+=c^+\frac{\Gamma_-}{D}\Gamma_{\varphi}^2\sin2\theta\; ,
\end{equation}

\begin{equation}\label{c-}
    c_-=-ic^+\frac{\Gamma_-}{D}\Omega_R\Gamma_{\varphi}\sin2\theta\;
    ,
\end{equation}

\begin{equation}\label{Det}
    D=\frac{\Gamma_{\varphi}^2+\delta^2}{T_1}+\Gamma_{\varphi}\Omega_1^2\;
    .
\end{equation}
By choosing $c^+=1$ and using Eqs.~(\ref{sigma-x-a}) and
(\ref{sigma-y-a}) to calculate (\ref{mxy}) we find
\begin{equation}\label{m_x}
    M_x=M_c\cos\omega_{\rm mw} t+M_s\sin\omega_{\rm mw} t\; ,
\end{equation}
\begin{equation}\label{m_y}
    M_y=-M_s\cos\omega_{{\rm mw}} t+M_c\sin\omega_{{\rm mw}} t\; ,
\end{equation}
where
\begin{equation}\label{mc}
    M_c=M_0\frac{T_2^2\delta\gamma B_1}{1+\left(T_2\delta\right)^2+T_1T_2\left(\gamma
    B_1\right)^2}\; ,
\end{equation}

\begin{equation}\label{ms}
    M_s=M_0\frac{T_2\gamma B_1}{1+\left(T_2\delta\right)^2+T_1T_2\left(\gamma
    B_1\right)^2}\; .
\end{equation}
Therefore, the steady-state solution of the density matrix
equations (\ref{ro})-(\ref{ro-}) and (\ref{k})-(\ref{kap-})
provides the well known expressions for the longitudinal
(\ref{mzst}) and transverse components (\ref{mc}), (\ref{ms}) of
the magnetization \cite{Abr}.

\section{Low-frequency response of an irradiated two-level
system}

The response of a two-level system to a weak external force
$f(t)$, in the limit of vanishing force $f(t)$, is found from the
following equations:

\begin{multline}\label{roLFnmr}
    \frac{d\rho}{dt}=-A_1\rho+B\rho_+-
    \frac{i}{\hbar} f(t)\rho_-\sin2\theta+(\Gamma_-)\cos2\theta\;
    ,
\end{multline}

\begin{multline}\label{roLF2nmr}
    \frac{d\rho_+}{dt}=-i\Omega_R\rho_-+B\rho-A_2\rho_++\frac{i}{\hbar} f(t)\rho_-\cos2\theta\\
    +(\Gamma_-)\sin2\theta\; ,
\end{multline}

\begin{multline}\label{roLF3nmr}
     \frac{d\rho_-}{dt}=-i\Omega_R\rho_+-\Gamma_{\varphi}\rho_-+
     \frac{i}{\hbar}
     f(t)\left(\rho_+\cos2\theta-\rho\sin2\theta\right)\; .
\end{multline}
By taking the functional derivative with respect to $f(t')$ we
obtain
\begin{multline}\label{fdro}
    \left(\frac{d}{dt}+A_1\right)\frac{\delta\rho(t)}{\delta f(t')}-
    B\frac{\delta\rho_+(t)}{\delta f(t')}=
    -\frac{i}{\hbar}\delta(t-t')\rho_-(t')\sin2\theta\,
\end{multline}
\begin{multline}\label{fdro+}
    \left(\frac{d}{dt}+A_2\right)\frac{\delta\rho_+(t)}{\delta f(t')}-
    B\frac{\delta\rho(t)}{\delta f(t')}+i\Omega_R\frac{\delta\rho_-(t)}{\delta
    f(t')}\\=
    \frac{i}{\hbar}\delta(t-t')\rho_-(t')\cos2\theta\; ,
\end{multline}
\begin{multline}\label{fdro-}
    \left(\frac{d}{dt}+\Gamma_{\varphi}\right)\frac{\delta\rho_-(t)}{\delta f(t')}
    +i\Omega_R\frac{\delta\rho_+(t)}{\delta
    f(t')}\\=
    \frac{i}{\hbar}\delta(t-t')[\rho_+(t')\cos2\theta-\rho(t')\sin2\theta]\;
    .
\end{multline}
where we used the definition $\delta f(t)/\delta
f(t')=\delta(t-t')$.

The corresponding susceptibilities $\chi_{\rho}(\omega)$,
$\chi_{\rho_+}(\omega)$, $\chi_{\rho_-}(\omega)$ are defined
similar to Eq.~(\ref{chi}):
\begin{equation}\label{fdchi}
\frac{\delta\rho(t)}{\delta
f(t')}=\int\frac{d\omega}{2\pi}\exp^{-i\omega(t-t')}\chi_{\rho}(\omega),\
\text{etc.}
\end{equation}

From Eqs.~(\ref{fdro}),  (\ref{fdro+}), and (\ref{fdro-}) we can
readily find the susceptibilities $\chi_{\rho}(\omega)$,
$\chi_{\rho_+}(\omega)$, and $\chi_{\rho_-}(\omega)$:
\begin{widetext}
\begin{equation}\label{hiro}
    \chi_{\rho}(\omega)=-\frac{i}{\hbar D(\omega)}\left\{
   \rho_-(t')\sin
   2\theta\left[\left(-i\omega+\Gamma_{\varphi}\right)\left(-i\omega+\frac{1}{T_1}\right)+
   \Omega_R^2\right]-\Omega_R B\Big[\rho_+(t')\cos 2\theta-\rho(t')\sin
   2\theta\Big]\right\}\; ,
\end{equation}

\begin{eqnarray}\label{hiro+}
    \chi_{\rho_+}(\omega)=\frac{i}{\hbar D(\omega)}\Big\{
   \rho_-(t')\cos
   2\theta\left(-i\omega+\Gamma_{\varphi}\right)\left(-i\omega+A_1\right)
-\rho_-(t')\sin 2\theta B\left(-i\omega+\Gamma_{\varphi}\right)
\\-
    i\Omega_R \Big[\rho_+(t')\cos 2\theta-\rho(t')\sin
   2\theta\Big]\left(-i\omega+A_1\right)\Big\}\nonumber\;
   ,
\end{eqnarray}

\begin{equation}\label{hiro-}
    \chi_{\rho_-}(\omega)=\frac{i}{\hbar D(\omega)}\left(-i\omega+\frac{1}{T_1}\right)
    \Big\{\left(-i\omega+\Gamma_{\varphi}\right)
\Big[\rho_+(t')\cos 2\theta-\rho(t')\sin
2\theta\Big]-i\Omega_R\rho_-(t')\cos 2\theta\Big\}\; ,
\end{equation}

\end{widetext}
where
\begin{equation}\label{Domega}
    D(\omega)=\left(-i\omega+\Gamma_{\varphi}\right)^2
    \left(-i\omega+\frac{1}{T_1}\right)+\left(-i\omega+A_1\right)\Omega_R^2\;
    .
\end{equation}
The functional derivatives (\ref{fdchi}) are defined for $t>t'$,
where $t'$ is the time the external force is being applied. Hence,
the susceptibilities (\ref{hiro}), (\ref{hiro+}), and
(\ref{hiro-}) describe the evolution of the system for times
$t>t'$, with the $\rho(t')$'s, corresponding to when the external
force is applied. Therefore, the subsequent evolution of the
system depends on its state just before the perturbation is
applied.

In what follows we consider two cases. The first one is when the
relaxation time $T_1$ is relatively short. In this case the system
quickly reaches thermal equilibrium during the measurement. For
this case, we take for $\rho(t')$, $\rho_+(t')$, and $\rho_-(t')$,
their steady-state values: $\rho^{(0)}$ from Eq.~(\ref{st0}),
$\rho_+^{(0)}$ from Eq.~(\ref{st+}), and $\rho_-^{(0)}$ from
Eq.~(\ref{st-}). For the corresponding susceptibilities, we obtain
the following expressions\cite{Green1}:

\begin{widetext}
\begin{equation}\label{hiro1}
    \chi_{\rho}(\omega)=-\frac{\Omega_R}{\hbar D(\omega)\Gamma_{\varphi}}
   \rho_+^{(0)}\left\{\sin2\theta\left[\left(-i\omega+\Gamma_{\varphi}\right)
   \left(-i\omega+\frac{1}{T_1}\right)+\Omega^2_R\right]
   +\frac{\Omega_R^2}{\Gamma_{\varphi}}B\cos2\theta\right\}\; ,
\end{equation}

\begin{equation}\label{hiro+1}
\chi_{\rho_+}(\omega)=\frac{\Omega_R}{\hbar
D(\omega)\Gamma_{\varphi}}\cos2\theta
  \rho_+^{(0)}\left[\left(-i\omega+\Gamma_{\varphi}\right)\left(-i\omega+\frac{1}{T_1}\right)
  -\left(-i\omega+A_1\right)\frac{\Omega_R^2}{\Gamma_{\varphi}}\right]\;
  ,
\end{equation}

\begin{equation}\label{hiro-1}
\chi_{\rho_-}(\omega)=i\frac{\Omega_R^2}{\hbar
D(\omega)\Gamma_{\varphi}^2}\rho_+^{(0)}\cos2\theta
\left(-i\omega+\frac{1}{T_1}\right)\left(-i\omega+2\Gamma_{\varphi}\right)\;
.
\end{equation}
\end{widetext}
The other case is when the spin-lattice relaxation time $T_1$ is
extremely long compared to the measurement time. If we measure the
spin noise in this case, then it is reasonable to take $\rho{(0)}=
-1/\sqrt{N}$, $\rho{(0)}_+=\rho{(0)}_-=0$ (see the explanation in
Section III). For the corresponding susceptibilities, we obtain
 from (\ref{hiro}), (\ref{hiro+}), and
(\ref{hiro-}) (in the limit $1/T_1\ll\omega$):
\begin{equation}\label{hiro2}
    \chi_{\rho}(\omega)=-\frac{i\delta\Omega_1^2\Gamma_{\varphi}}{\hbar\sqrt{N}
    d(\omega)\Omega_R^2}\; ,
\end{equation}

\begin{equation}\label{hiro+2}
    \chi_{\rho_+}(\omega)=\frac{\Omega_1}{\hbar\sqrt{N}
    d(\omega)}\left(-i\omega+\Gamma_{\varphi}\frac{\Omega^2_1}{\Omega^2_R}\right)\;
    ,
\end{equation}

\begin{equation}\label{hiro-2}
    \chi_{\rho_-}(\omega)=\frac{\omega\Omega_1}{\hbar\sqrt{N}
    d(\omega)\Omega_R}\left(-i\omega+\Gamma_{\varphi}\right)\; ,
\end{equation}
where
\begin{equation}\label{dom}
    d(\omega)=\Gamma_{\varphi}\left(\Omega_1^2-2\omega^2\right)
    +i\omega\left(\omega^2-\Omega_R^2-\Gamma_{\varphi}^2\right)\;
    .
\end{equation}


\begin{thebibliography}{99}
\bibitem{Comm1} The substitution of the actual photon number
$N_{ph}$ by its average value $\langle N_{ph}\rangle$ means that
we treat the driving field classical. In this sence the Eqs.
\ref{ro}, \ref{ro+}, and \ref{ro-} describe the evolution of Rabi
levels originated from the interaction of two-level system with a
classical (coherent state) driving field.
\end{thebibliography}
\end{document}